\documentclass[a4paper,12pt]{article}
\usepackage{amssymb}
\usepackage{mathrsfs}
\usepackage{bbm}
\usepackage{epsf,amsmath,amssymb}
\usepackage[dvips,usenames]{color}
\usepackage{graphicx}
\usepackage{color}
\usepackage{colortbl}
\usepackage{epsfig}

\newlength{\dinwidth}
\newlength{\dinmargin}
\def \eff{{\text{eff}}}

\def\c{C}
\def\cs{{\c_7}}
\def\cn{{\c_9}}
\def\ct{{\c_{10}^{\rm eff}}}
\def\cne{\cn^{\rm eff}}
\def\cse{\cs^{\rm eff}}

\def\gl{\Gamma}

\def\l{\ell}
\def\ks{{K^{\ast}}}

\def\d{{\rm d}}

\def\mh{\hat{m}}
\def\mbh{\mh_b}
\def\mph{\mh_K}
\def\mvh{\mh_{K^*}}

\def\mlh{\mh_\l}

\def\sh{\hat{s}}

\def\a{{\cal A}}

\def\uh{{\hat{u}}}
\def\la{{\lambda}}

\def\be{\begin{equation}}
\def\ee{\end{equation}}
\def\ba{\begin{eqnarray}}
\def\ea{\end{eqnarray}}

\begin{document}

\title{\bf $B \to \ks \ell^+ \ell^-$, $K \ell^+ \ell^-$ decays in a family non-universal $Z^{\prime}$ model}
\author{Qin Chang$^{a,b}$, Xin-Qiang
Li$^{a,c}$$\footnote{Corresponding author}$, Ya-Dong
Yang$^{b,d}$\\
{ $^a$\small Department of Physics, Henan Normal University,
Xinxiang, Henan 453007, P.~R. China}\\
{ $^b$\small Institute of Particle Physics, Huazhong Normal
University, Wuhan, Hubei  430079, P.~R. China}\\
{ $^c$\small Institut f\"ur Theoretische Physik E, RWTH Aachen
University, D--52056 Aachen, Germany}\\
{ $^d$\small Key Laboratory of Quark \& Lepton Physics, Ministry of
Education, P.~R. China}}
\date{}
\maketitle
\bigskip\bigskip
\maketitle \vspace{-1.5cm}

\begin{abstract}
Motivated by the observed forward-backward asymmetry in $B\to K^{\ast} \ell^+\ell^-$ decay, we perform a detailed analysis of this decay mode within a family non-universal $Z^{\prime}$ model. With the related coupling $Z^{\prime}-\bar{s}b$ constrained by $B_s-\bar{B}_s$ mixing, $B\to \pi K$ and $B\to X_s \mu^{+}\mu^{-}$ decays, we look for further constraint on the couplings $Z^{\prime}-\mu^{+}\mu^{-}$ from $A_{FB}(B\to K^{\ast} \mu^{+}\mu^{-})_{0 {\rm GeV}^2\leqslant q^2\leqslant 2 {\rm GeV}^2}$ and get numerically $B_{\mu\mu}^{L,R}\sim{\cal O}(10^{-2})$. Moreover, we find that the relations $B_{\mu\mu}^{L}<B_{\mu\mu}^{R}$ and $B_{\mu\mu}^{L}+B_{\mu\mu}^{R}<0$, with a small negative phase $\phi_{s}^{L}$, are crucial to moderate the discrepancy for $A_{FB}(B\to K^{\ast} \mu^{+}\mu^{-})$ between the SM prediction and the experimental data. Numerically, comparing with the SM prediction, we find that $A_{FB}(B\to K^{\ast} \mu^{+}\mu^{-})_{0 {\rm GeV}^2\leqslant q^2\leqslant 2 {\rm GeV}^2}$ could be enhanced about $80\%$ and $50\%$ by $Z^{\prime}$ contribution at most in scenarios S1 and S2, corresponding to the two fitted results of $\phi_{s}$ by UTfit collaboration, respectively. However, the
results are still about $1.5\sigma$ lower than the experimental measurement.
\end{abstract}

\noindent{{\bf Keywords:} B-physics, Rare decays, Beyond Standard Model}

\noindent{{\bf ArXiv ePrint:} 1002.2758}

\newpage

\section{Introduction}

As is well-known, the electro-weak penguin decays $b\to s \ell^+\ell^-$ appear only at the one-loop level in the Standard Model~(SM), and are therefore very sensitive to possible new physics~(NP) beyond the SM.
Among many inclusive and exclusive processes based on the quark level $b\to s \ell^+\ell^-$ transition, the exclusive $B\to K^* \ell^+\ell^-$ decays are of particular interest in this respect, as many observables in these decays, such as the branching ratio, the longitudinal polarization fraction, the forward-backward asymmetry $A_{FB}$, and the isospin asymmetry, could be used to test the SM and to probe possible NP.

In the literature, the exclusive $B \to K^* \ell^+ \ell^-$ decays have been investigated in great detail by many authors, both in the SM and within various NP models~\cite{SM_ANA,Burdman:1998mk,Beneke:2001at,Ali:1999mm,Altmannshofer:2008dz,NP_ANA}. Among these observables of $B \to K^* \ell^+ \ell^-$, the $A_{FB}$ is particularly useful. As discussed in Refs.~\cite{Burdman:1998mk,Beneke:2001at,Ali:1999mm}, the zero of $A_{FB}$ is largely free from hadronic uncertainties in the SM and hence could be a powerful probe for various NP models. Recently this observable has been measured as a function of the dilepton invariant mass square $q^2=M^2_{ll}$, by both BaBar~\cite{:2008ju,Aubert:2008ps} and Belle~\cite{Ishikawa:2006fh,:2009zv} collaborations. Their fitted
$A_{FB}$ spectrum is generally higher than the SM expectation in all $q^2$ bins. Especially, the recent measurement from Belle collaboration~\cite{:2009zv} shows
\begin{equation}
A_{FB}(B \to K^* \ell^+ \ell^-)_{0 {\rm GeV}^2 \leqslant q^2\leqslant2
{\rm GeV}^2}=0.47^{+0.26}_{-0.32}\pm0.03,
\end{equation}
favoring a positive value, whereas the sign of the SM prediction for $A_{FB}(B \to K^* \ell^+ \ell^-)\sim-0.1$ at $0 {\rm GeV}^2\leqslant q^{2}\leqslant 2 {\rm GeV}^2 $ is negative. Such a large discrepancy is hard to be moderated within the SM. Furthermore, the measurements prefer positive values for $A_{FB}(B \to K^* \ell^+ \ell^-)$ in the whole $q^2$ region, indicating that there might be no zero crossing, which is apparently contrary to the SM prediction~\cite{Burdman:1998mk,Beneke:2001at,Ali:1999mm}.

The measurements have motivated many recent investigations on the possible mismatch~\cite{Chiang,Chen:2009bj,Alok:2009tz}. In this paper, we revisit this decay mode within a family non-universal $Z^{\prime}$ model~\cite{Langacker}, which could be naturally derived in certain string constructions, $E_6$ models and so on. In our previous paper~\cite{Yang1}, with the constraints from $B_s-\bar{B}_s$ mixing, $B\to \pi K$ and $B\to X_s\mu^{+}\mu^{-}$ decays, we have obtained an explicit picture for the $Z^{\prime}$ couplings $B_{sb}^{L}$ and $B_{\mu\mu}^{L,R}$. Thus, it is of interest to see whether the discrepancy of $A_{FB}(B \to K^* \ell^+ \ell^-)$ at $0\leqslant q^2\leqslant2 {\rm GeV}^2$ between the SM prediction and the experimental data could be moderated with the constrained non-universal $Z^{\prime}$ couplings.

Our paper is organized as follows. In Sec.~2, we present a brief review of the SM theoretical framework for $B \to K^{(*)} \ell^+ \ell^-$ decays. After brief introduction of the employed family non-universal $Z^{\prime}$ model in Sec.~3, we present our numerical analyses and discussions in Sec.~4. Our conclusions are summarized in Sec.~5. Appendices A and B include our theoretical inputs.

\section{The SM prediction}

Neglecting the doubly Cabibbo-suppressed contributions, the effective Hamiltonian governing $b\to s \ell^+\ell^-$ transition is given by~\cite{Altmannshofer:2008dz,Chetyrkin:1996vx}
\begin{equation} \label{eq:Heff}
{\cal H}_{\eff} = - \frac{4\,G_F}{\sqrt{2}}V_{tb}V_{ts}^{\ast}
\sum_{i=1}^{10} C_i(\mu) O_i(\mu) \,,
\end{equation}
where explicit expressions of $O_{i}$ could be found in Ref.~\cite{Altmannshofer:2008dz}, and the Wilson coefficients $C_i$ can be calculated perturbatively~\cite{Beneke:2001at,bobeth,bobeth02,Huber:2005ig}, with the numerical results listed in Table~\ref{wc}.

\begin{table}[htbp]
 \begin{center}
 \caption{The SM Wilson coefficients at the scale $\mu=m_b$.}
 \label{wc}
 \vspace{0.3cm}
 \doublerulesep 0.5pt \tabcolsep 0.07in
 \begin{tabular}{lccccccccccc}
 \hline \hline
 $C_1(m_b)$& $C_2(m_b)$& $C_3(m_b)$& $C_4(m_b)$& $C_5(m_b)$& $C_6(m_b)$& $C_7^{\rm eff}(m_b)$& $C_9^{\rm eff}(m_b)-Y(q^2)$& $C_{10}^{\rm eff}(m_b)$\\\hline
 $-0.274$  & $1.007$   & $-0.004$  & $0.076$   & $0.000$   & $0.001$   & $-0.302$            & $4.094$   & $-4.193$\\
  \hline \hline
 \end{tabular}
 \end{center}
 \end{table}

The effective coefficients $C_{7,9,10}^{{\rm eff}}$ in Table~\ref{wc} are defined respectively as~\cite{Buras:1993xp}
\begin{eqnarray}\label{eq:effWC}
&& C_7^{\rm eff} = \frac{4\pi}{\alpha_s}\, C_7 -\frac{1}{3}\, C_3 -
\frac{4}{9}\, C_4 - \frac{20}{3}\, C_5\, -\frac{80}{9}\,C_6\,,
\nonumber\\
&& C_9^{\rm eff} = \frac{4\pi}{\alpha_s}\,C_9 + Y(q^2)\,, \qquad
   C_{10}^{\rm eff} = \frac{4\pi}{\alpha_s}\,C_{10}\,,
\end{eqnarray}
where $Y(q^2)$ denotes the matrix element of four-quark operators and is known from the literature~\cite{Ali:1999mm,Altmannshofer:2008dz,Ali:1991is,Buras:1994dj}. We have neglected the long-distance contribution mainly due to $J/\Psi$ and $\Psi^{\prime}$ in the decay chain $B\to K^{(*)}\Psi^{(\prime)} \to K^{(*)}\ell^{+}\ell^{-}$, which could be vetoed experimentally~\cite{Aubert:2008ps,:2009zv}. For recent detailed discussion of such resonance effects, we refer to Ref.~\cite{resEff}.

Although there are quite a lot of interesting observables in $B\to K^{(*)} \ell^+ \ell^-$ decay, in this paper we shall focus only on the dilepton invariant mass spectrum and the forward-backward asymmetry.

Adopting the same convention and notation as \cite{Ali:1999mm}, the dilepton invariant mass spectrum for $B\to K^{(*)}\ell^+\ell^-$ decay is given respectively as~\cite{Ali:1999mm,Geng:1996az}
\begin{eqnarray}
\frac{\d \gl^{K}}{\d\sh}& = & \frac{G_F^2 \, \alpha^2 \, m_B^5}{2^{10} \pi^5}
      \left| V_{ts}^\ast  V_{tb} \right|^2 \, \uh(\sh) \,
      \Bigg\{(|A^{\prime}|^2+|C^{\prime}|^2)\Big(\lambda-\frac{\uh(\sh)^2}{3}\Big)
      +|C^{\prime}|^24\mlh^2(2+2\mph^{2}-\sh)\, \nonumber\\
      & & + {\rm Re}(C^{\prime}D^{\prime\ast})8\mlh^2(1-\mph^2)+|D^{\prime}|^24\mlh^2\sh
      \Bigg\}\,, \\ [0.2cm]
\frac{\d \gl^{K^*}}{\d\sh} & = &
  \frac{G_F^2 \, \alpha^2 \, m_B^5}{2^{10} \pi^5}
      \left| V_{ts}^\ast  V_{tb} \right|^2 \, \uh(\sh) \,
      \Bigg\{
\frac{|A|^2}{3} \sh \la (1+2 \frac{\mlh^2}{\sh}) +|E|^2 \sh
\frac{\uh(\sh)^2}{3}
        \Bigg.
        \nonumber \\
  & & + \Bigg. \frac{1}{4 \mvh^2} \left[
|B|^2 (\la-\frac{\uh(\sh)^2}{3} + 8 \mvh^2 (\sh+ 2 \mlh^2) )
          + |F|^2 (\la -\frac{ \uh(\sh)^2}{3} + 8 \mvh^2 (\sh- 4 \mlh^2))
\right]
        \Bigg.
        \nonumber \\
  & & +\Bigg.
   \frac{\la }{4 \mvh^2} \left[ |C|^2 (\la - \frac{\uh(\sh)^2}{3})
 + |G|^2 \left(\la -\frac{\uh(\sh)^2}{3}+4 \mlh^2(2+2 \mvh^2-\sh) \right)
\right]
        \Bigg.
        \nonumber \\
  & & - \Bigg.
   \frac{1}{2 \mvh^2}
\left[ {\rm Re}(BC^\ast) (\la -\frac{ \uh(\sh)^2}{3})(1 - \mvh^2 -
\sh)
\nonumber  \right. \Bigg.\\
& & + \left.  \Bigg.
       {\rm Re}(FG^\ast) ((\la -\frac{ \uh(\sh)^2}{3})(1 - \mvh^2 - \sh) +
4 \mlh^2 \la) \right]
        \Bigg.
        \nonumber \\
  & & - \Bigg.
 2 \frac{\mlh^2}{\mvh^2} \la  \left[ {\rm Re}(FH^\ast)-
 {\rm Re}(GH^\ast) (1-\mvh^2) \right] +\frac{\mlh^2}{\mvh^2} \sh \la |H|^2
  \Bigg\} \; .
   \label{eq:dwbvll}
\end{eqnarray}
Here the auxiliary functions $A^{(\prime)},...$, with the explicit expressions given in \cite{Ali:1999mm}, are combinations of the effective Wilson coefficients Eq.~(\ref{eq:effWC}) and the $B\to K^{(*)}$ transition form factors, which are calculated with light-cone QCD sum rule approach in Ref.~\cite{Ball:2004rg}.

The differential forward-backward asymmetry for $B\to K^*\ell^+\ell^-$ decay is defined  as~\cite{Ali:1999mm}
\begin{eqnarray}  \label{eq:dfbabvllex}
  \frac{\d \a_{\rm FB}^{K^{\ast}}}{\d \sh}& =&
  -\frac{G_F^2 \, \alpha^2 \, m_B^5}{2^{8} \pi^5}
      \left| V_{ts}^\ast  V_{tb} \right|^2 \, \sh \, \uh(\sh)^2 \, \nonumber \\
& & \hspace{-0.5cm} \times \left[  {\rm Re}(\cne\ct^{\ast}) V A_1+ \frac{\mbh}{\sh}
{\rm Re}(\cse\ct^{\ast}) {\Big (}V T_2 (1-\mvh)+ A_1 T_1
(1+\mvh){\Big)} \right].
\end{eqnarray}
It is noted that, although $\ct$ is real in the SM, it could become complex after including the $Z^{\prime}$ contributions. Note that the $A_{FB}$ for $B\to K \ell^+\ell^-$ decay vanish both in the SM and within the $Z^{\prime}$ model considered in the present paper, since neither further operators nor higher-order corrections are included in our discussion~\cite{Ali:1999mm,Bobeth:2001sq}. From the experimental point of view, the normalized forward-backward asymmetry is more useful, which is defined
as~\cite{Ali:1999mm}
\begin{equation} \label{eq:AFB2}
  \frac{\d \bar{\a}_{\rm FB}}{\d \sh} =
\frac{\d \a_{\rm FB}}{\d \sh}/\frac{\d \gl}{\d\sh}\,.
\end{equation}

\section{Family non-universal $Z^{\prime}$ couplings and their effects }

A family non-universal $Z^{\prime}$ model, which has been formulated in detail by Langacker and Pl\"{u}macher~\cite{Langacker}, can lead to FCNC processes even at tree level due to the non-diagonal chiral coupling matrix. Under the assumption that the couplings of right-handed quark flavors with $Z^{\prime}$ boson are diagonal, the $Z^{\prime}$ part of the effective Hamiltonian for $b\to s l^+ l^-$ can be written as~\cite{Yang1,Liu}
\begin{equation}\label{ZPHbsll}
 {\cal H}_{eff}^{Z^{\prime}}(b\to sl^+l^-)=-\frac{2G_F}{\sqrt{2}}
 V_{tb}V^{\ast}_{ts}\Big[-\frac{B_{sb}^{L}B_{ll}^{L}}{V_{tb}V^{\ast}_{ts}}
 (\bar{s}b)_{V-A}(\bar{l}l)_{V-A}-\frac{B_{sb}^{L}B_{ll}^{R}}{V_{tb}V^{\ast}_{ts}}
 (\bar{s}b)_{V-A}(\bar{l}l)_{V+A}\Big]+{\rm h.c.}\,,
\end{equation}
which could also be reformulated as
\begin{equation}\label{ZPHbsllMo}
 {\cal H}_{eff}^{Z^{\prime}}(b\to sl^+l^-)=-\frac{4G_F}{\sqrt{2}}
 V_{tb}V^{\ast}_{ts}\left[\triangle C_9^{\prime} O_9+\triangle C_{10}^{\prime} O_{10}\right]+{\rm h.c.}\,,
\end{equation}
with
\begin{eqnarray}\label{C910Zp}
 \triangle C_9^{\prime}(M_W)&=&-\frac{g_s^2}{e^2}\frac{B_{sb}^L
 }{V_{ts}^{\ast}V_{tb}} (B_{ll}^{L}+B_{ll}^{R})\,,\nonumber \\
 \triangle C_{10}^{\prime}(M_W)&=&\frac{g_s^2}{e^2}\frac{B_{sb}^L
 }{V_{ts}^{\ast}V_{tb}} (B_{ll}^{L}-B_{ll}^{R}),
\end{eqnarray}
where $B_{sb}^L$ and $B_{ll}^{L,R}$ denote the effective chiral $Z^{\prime}$ couplings to quarks and leptons.

With the above expressions, the $Z^{\prime}$ contributions can be represented as modifications of the Wilson coefficient of the corresponding semileptonic operators, i.e., $C^{\prime}_{9,10}(M_W)=C_{9,10}^{SM}(M_W)+\triangle C_{9,10}^{\prime}(M_W)$. The running from $M_{W}$ scale down to $m_{b}$ is the same as the SM ones\cite{Altmannshofer:2008dz, Chetyrkin:1996vx}. Numerically, with the central value of the inputs, we get
\begin{eqnarray}
\label{NuVaC9}
 C_9^{\prime}(m_b)&=&0.0682-28.82\frac{B_{sb}^L
 }{V_{ts}^{\ast}V_{tb}} (B_{ll}^{L}+B_{ll}^{R})\,,\\
 \label{NuVaC0}
 C_{10}^{\prime}(m_b)&=&-0.0695+28.82\frac{B_{sb}^L
 }{V_{ts}^{\ast}V_{tb}} (B_{ll}^{L}-B_{ll}^{R}),
\end{eqnarray}

To include $Z^{\prime}$ effects, one just needs to make the replacements
\begin{eqnarray}\label{C910ERp}
C_{9}^{\rm eff}&\to&\bar{C}_9^{\rm eff}=\frac{4\pi}{\alpha_s}C_9^{\prime}+Y(q^2)\;,\nonumber\\
C_{10}^{\rm eff}&\to&\bar{C}_{10}^{\rm eff}=\frac{4\pi}{\alpha_s}C_{10}^{\prime}\;.
\end{eqnarray}
in the formalisms relevant to $B\to K^{(\ast)} \ell^{+}\ell^{-}$ listed in section 2. With the formulae collected above,  we shall  proceed to present our numerical analyses and discussions in the next section.

\section{Numerical analyses and discussions}

\begin{figure}[t]
\begin{center}
\epsfxsize=15cm \centerline{\epsffile{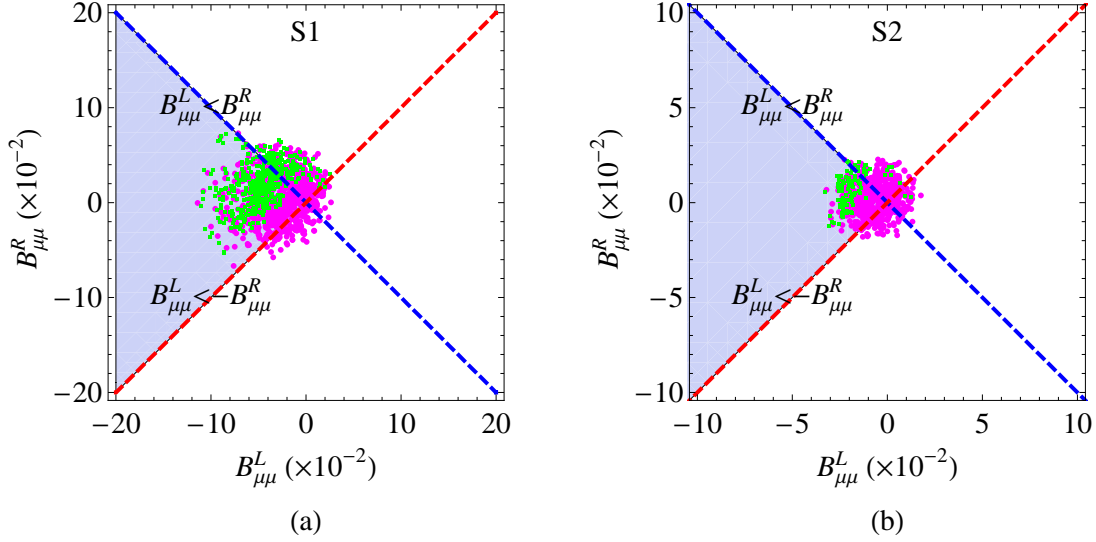}}
\centerline{\parbox{16cm}{\caption{\label{FigBmumuLR}\small The allowed regions for the parameters $B^{L,R}_{\mu\mu}$. The pink regions are allowed by the constraints from $B\to X_s\mu^{+}\mu^{-}$ decay~\cite{Yang1}. The green ones correspond to the final parameter space with the constraint from $A_{FB}(B\to K^{\ast} \mu^{+}\mu^{-})_{0 {\rm GeV}^2\leqslant q^2\leqslant 2 {\rm GeV}^2}$ also included. The shaded region is derived from our qualitative analysis with the conditions  $B_{ll}^{L}+B_{ll}^{R}<0$ and $B_{ll}^{L}<B_{ll}^{R}$.}}}
\end{center}
\end{figure}

The considered $Z^{\prime}$ contributions to $B\to K\mu^{+}\mu^{-}$ and $B\to K^{*}\mu^{+}\mu^{-}$ decays involve three couplings: $B^{L}_{sb}$, $B^{L}_{\mu\mu}$ and $B^{R}_{\mu\mu}$. In our previous papers~\cite{Yang1}, we have combined $B_s-\bar{B}_{s}$ mixing, $B\to\pi K^{(\ast)}$ and $\rho K$,
$B\to X_s\mu^{+}\mu^{-}$, as well as $B_s\to\mu^+\mu^-$ decays to constrain these $Z^{\prime}$ couplings and their possible phase. Our combined results are re-tabulated in Table~\ref{NPPara_value} and re-displayed in Fig.~\ref{FigBmumuLR}~(pink region). The two solutions S1 and S2 for $|B_{sb}^{L}|$ and $\phi_{s}^L$ correspond to the two fitted results for the new physics parameter $\phi_{B_s}$ performed by the UTfit collaboration~\cite{UTfit}.  It is natural to question whether the constrained parameter space could account for the $A_{FB}$ measured recently by the Belle collaboration~\cite{:2009zv}.

The $\mathcal{A}_{FB}(s)$ spectrum for $B\to K^{\ast} \mu^{+}\mu^{-}$ decay measured by Belle collaboration tends to be shifted toward the positive side in all six $q^2$ bins, indicating
that there might be no zero crossing. However, a zero crossing in $\mathcal{A}_{FB}(s)$, whose position is well-determined and free from hadronic uncertainties at the leading order in $\alpha_s$, is well predicted in the SM~\cite{Burdman:1998mk,Beneke:2001at,Ali:1999mm}. Especially in the bin $0 {\rm GeV}^2< q^2\leqslant 2 {\rm GeV}^2$, the sign of the SM prediction is negative, being different from the experimental measurement, which favors a positive value on the other hand.

Comparing the two terms in the square bracket in Eq.~(\ref{eq:dfbabvllex}), one can see that, at low $q^2$ region the first term ${\rm Re}(\cne\ct^{\ast})$ is suppressed by one power of $q^2/m_b^2$ relative to the second one, and $A_{FB}(B\to K^{\ast}\mu^{+}\mu^{-})$ is therefore dominated by the second term ${\rm Re}(\cse\ct^{\ast})$. Thus, at low $q^2$ region $A_{FB}(B\to K^{\ast}\mu^{+}\mu^{-})$ could be significantly changed by flipping the sign of ${\rm Re}(\cse\ct^{\ast})$ from a positive in the SM~(due to $C_7^{\rm eff}<0$ and $C_{10}^{\rm eff}<0$) to a negative one. On the other hand, in order to keep
$A_{FB}(B\to K^{\ast}\mu^{+}\mu^{-})$ positive in the high $q^2$ region, the sign of
${\rm Re}(\cne\ct^{\ast})$ should be maintained negative as predicted in the SM.

With $\phi_s^L\sim-72^{\circ}$~(S1) and $\phi_s^L\sim-82^{\circ}$~(S2) obtained in Ref.~\cite{Yang1}, and keeping in mind that the CKM element $V_{ts}$ is negative, one can easily find from Eq.~(\ref{NuVaC0}) that the sign of ${\rm Re}(\cse\bar{C}_{10}^{{\rm eff}\ast})$ could be flipped if $B_{ll}^{L}<B_{ll}^{R}$. In order to see the $Z^{\prime}$ effect on ${\rm Re}(\bar{C}_9^{\rm eff}\bar{C}_{10}^{{\rm eff}\ast})$ explicitly, with $Y(q^2)$ excluded, we can
rewrite it as
\begin{eqnarray}
{\rm Re}(\bar{C}_9^{\rm eff}\bar{C}_{10}^{{\rm eff}\ast})&=&{\rm Re}(\bar{C}_9^{\rm eff})\,{\rm Re}(\bar{C}_{10}^{{\rm eff}\ast})+{\rm Im}(\bar{C}_9^{\rm eff})\,{\rm Im}(\bar{C}_{10}^{\rm eff})\,,\nonumber\\[0.2cm]
&\simeq&{\rm Re}(\bar{C}_9^{\rm eff})\,{\rm Re}(\bar{C}_{10}^{{\rm eff}\ast})+\left(\frac{4\pi}{\alpha_s}\right)^2\,{\rm Im}(\triangle
C_9^{\prime})\,{\rm Im}(\triangle C_{10}^{\prime})\,,
\end{eqnarray}
where the fact that both $C_{9}^{\rm eff}$ and $C_{10}^{\rm eff}$ are real in the SM has been used in the second line. To keep the sign of ${\rm Re}(\bar{C}_9^{\rm eff}){\rm Re}(\bar{C}_{10}^{{\rm eff}\ast})$ negative, one can derive the relation $B_{ll}^{L}+B_{ll}^{R}<0$ from Eq.~(\ref{NuVaC9}). At the same time, with the obtained relations, $B_{ll}^{L}+B_{ll}^{R}<0$ and $B_{ll}^{L}<B_{ll}^{R}$, the term ${\rm Im}(\triangle C_9^{\prime}){\rm Im}(\triangle C_{10}^{\prime})$ is automatically negative, and hence the sign of ${\rm Re}(\bar{C}_9^{eff}\bar{C}_{10}^{{\rm eff}\ast})$ is indeed maintained to be negative.
It is interesting to note that the allowed parameter space in Table~\ref{NPPara_value} constrained by $B\to X_s\mu^{+}\mu^{-}$ and $B_s\to\mu^{+}\mu^{-}$ decays~\cite{Yang1} satisfy the relations, $B_{ll}^{L}<B_{ll}^{R}$ and $B_{ll}^{L}+B_{ll}^{R}<0$. It is however unclear whether the parameter space could bridge the discrepancy of $A_{FB}(B\to K^{\ast}\mu^{+}\mu^{-})$ between the SM prediction and the experimental data. In the following numerical evaluation, we shall perform a fit combining the constraints from $B\to X_s\mu^{+}\mu^{-}$ and $A_{FB}(B\to
K^{\ast} \mu^{+}\mu^{-})_{0 {\rm GeV}^2\leqslant q^2\leqslant 2 {\rm GeV}^2}$, leaving the other observables for $B\to K^{(\ast)} \mu^{+}\mu^{-}$ decays as our predictions within such a $Z^{\prime}$ model.

For consistence, we take the same simplifications for the family non-universal $Z^{\prime}$ couplings as Ref.~\cite{Yang1}. Our fit is performed with the experimental data on $A_{FB}(B\to K^{\ast} \mu^{+}\mu^{-})_{0 {\rm GeV}^2\leqslant q^2\leqslant 2 {\rm GeV}^2}$ varying randomly within its $1.7\sigma$~($\simeq 90\%$~C.L.) error bar, while the theoretical uncertainties are obtained by varying the input parameters within their respective regions specified in Appendices A and B. Our re-fitted numerical results for $B_{\mu\mu}^{L,R}$ are listed in the columns 6-7 in Table~\ref{NPPara_value}, and the corresponding allowed regions are shown in Fig.~\ref{FigBmumuLR}~(green region). As illustrated in Fig.~\ref{FigBmumuLR}, after including the constraint from $A_{FB}(B\to K^{\ast} \mu^{+}\mu^{-})_{0 {\rm GeV}^2\leqslant q^2\leqslant 2 {\rm GeV}^2}$, the survived parameter spaces of $Z^{\prime}$ couplings~(green region) is further reduced. Comparing with the constraint from $B\to X_s \mu^{+}\mu^{-}$ decays~(pink region) only, one can see that the regions with $B_{\mu\mu}^{L}>B_{\mu\mu}^{R}$ are excluded, which confirms our naive analysis that the relation $B_{\mu\mu}^{L}<B_{\mu\mu}^{R}$ is needed to alleviate the observed discrepancy for $A_{FB}(B\to K^{\ast} \mu^{+}\mu^{-})_{0 {\rm GeV}^2\leqslant q^2\leqslant 2 {\rm GeV}^2}$. Numerically we also find that $B_{\mu\mu}^{L}<0$ and $|B_{\mu\mu}^{L}|>|B_{\mu\mu}^{R}|$, which means that the other condition $B_{\mu\mu}^{L}+B_{\mu\mu}^{R}<0$ is also kept.

\begin{table}[t]
 \begin{center}
 \caption{Columns 2-5 present the values for $B_{sb}^L$ with the constraints from $B_s-\bar{B}_{s}$ mixing and $B\to \pi K^{\ast}$, $\rho K$ decays and $B^{L,R}_{\mu\mu}$ with the constraints from $B\to X_s\mu^{+}\mu^{-}$ decay~\cite{Yang1}. Columns 6-7 are the re-fitted results after including the constraints from $B\to K^{\ast} \mu^+\mu^-$ decay. }
 \label{NPPara_value}
 \vspace{0.5cm}
 \doublerulesep 0.7pt \tabcolsep 0.1in
 \begin{tabular}{lccccccccccc} \hline \hline
  & $|B_{sb}^L|(\times10^{-3})$ & $\phi_{s}^L[^{\circ}]$ & $B^{L}_{\mu\mu}(\times10^{-2})$ & $B^{R}_{\mu\mu}(\times10^{-2})$& $B^{L}_{\mu\mu}(\times10^{-2})$ & $B^{R}_{\mu\mu}(\times10^{-2})$\\\hline
 S1 & $1.09\pm0.22$ & $-72\pm7$   & $-2.7\pm2.5$    & $0.61\pm2.4$  & $-4.75\pm2.44$    & $1.97\pm2.24$ \\
 S2 & $2.20\pm0.15$ & $-82\pm4$   & $-0.59\pm0.93$   & $0.19\pm0.88$& $-1.83\pm0.82$   & $0.68\pm0.85$ \\
  \hline \hline
 \end{tabular}
 \end{center}
 \end{table}

\begin{figure}[t]
\begin{center}
\epsfxsize=15cm \centerline{\epsffile{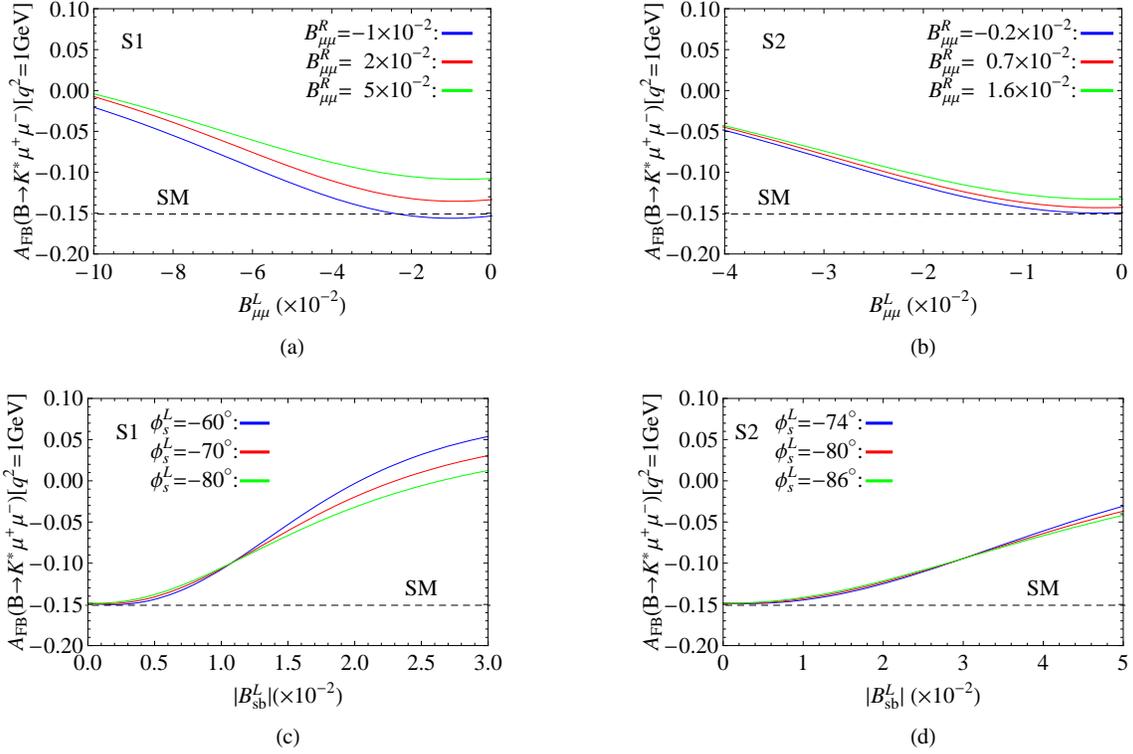}}
\centerline{\parbox{16cm}{\caption{\label{FigDBASp}\small The dependence of $A_{FB}(B\to K^{\ast}  \mu^+\mu^-)$ on $|B_{sb}^{L}|$, $\phi_s^{L}$ and $B_{\mu\mu}^{LR}$ with $q^2=1{\rm GeV}^2$. For comparison, the SM prediction is shown as dashed lines.}}}
\end{center}
\end{figure}

With the constrained $Z^{\prime}$ couplings and taking $q^2=1{\rm GeV}^2$, we show in Fig.~\ref{FigDBASp} the dependence of $A_{FB}(B\to K^{\ast} \mu^+\mu^-)$ on $|B_{sb}^{L}|$, $\phi_s^{L}$ and $B_{\mu\mu}^{L,R}$. From Fig.~\ref{FigDBASp}~(a) and (b), one can see that $A_{FB}(B\to K^{\ast} \mu^+\mu^-)$ at low $q^2$ bin could be enhanced to the experimental side by the $Z^{\prime}$ contribution with a large negative $B_{\mu\mu}^{L}$ and/or a positive $B_{\mu\mu}^{R}$. At the same time, from Fig.~\ref{FigDBASp}~(c) and (d) one can see that, with the constrained $|B_{sb}^{L}|$, a smaller phase $|\phi_{s}^{L}|$ is more helpful to enhance the $A_{FB}(B\to K^{\ast} \mu^+\mu^-)$ at low $q^2$ region. As a result, due to the fact that $|\phi_{s}^{L}|(S1)<|\phi_{s}^{L}|(S2)$, the solution S1 is preferable to S2. Moreover, a larger $|B_{sb}^{L}|$, which enlarges the $Z^{\prime}$ contribution, is also helpful for reconciling the $A_{FB}(B\to K^{\ast} \mu^+\mu^-)$ problem at low $q^2$.

\begin{figure}[t]
\begin{center}
\epsfxsize=15cm \centerline{\epsffile{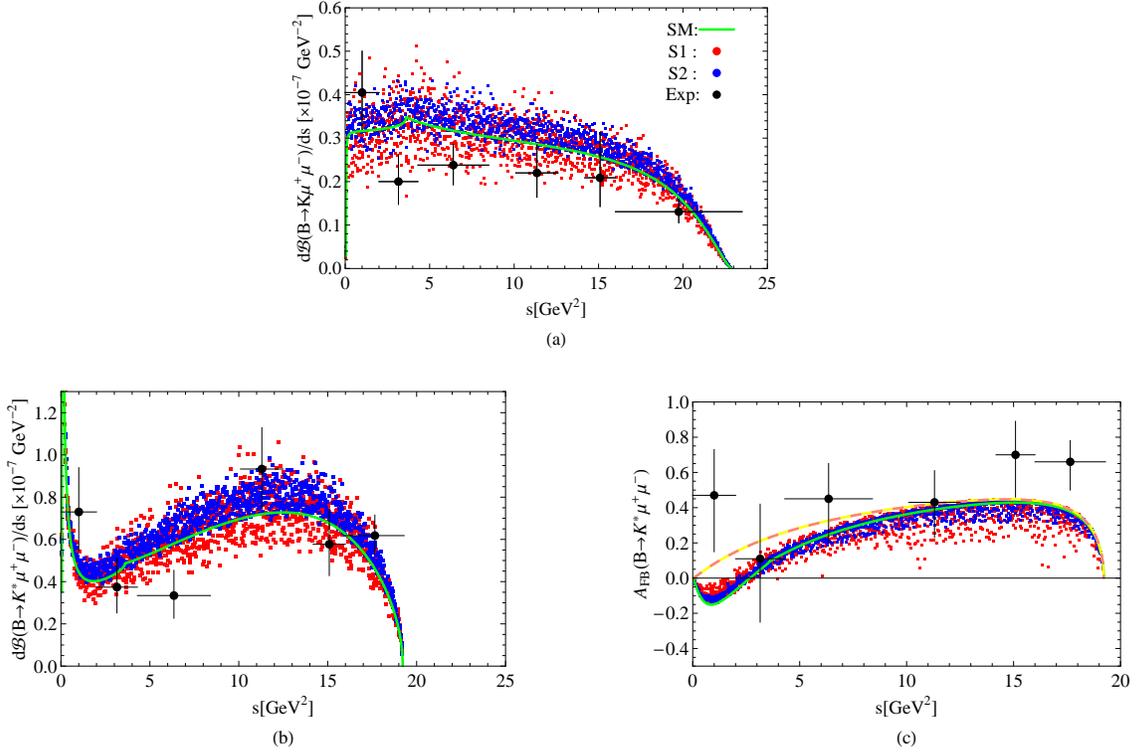}}
\centerline{\parbox{16cm}{\caption{\label{FigDBA}\small The effects
of the $Z^{\prime}$ contribution induced by $B_{\mu\mu}^{L,R}$ on
$d{\cal B}(B\to K^{(\ast)} \mu^+\mu^-)/d\hat{s}$ and $A_{FB}(B\to
K^{\ast} \mu^+\mu^-)$. The yellow (pink dashed) curve corresponds to
the special case with $B_{\mu\mu}^L=-0.3$ and $B_{\mu\mu}^R=0.01$ in
S1~(S2).}}}
\end{center}
\end{figure}

With the preferred choice $|B_{sb}^{L}|=1.1$~$(2.2)\times10^{-3}$, $\phi_s^{L}=-65^{\circ}$~$(-78^{\circ})$, $B_{\mu\mu}^{L}=-7.2$~$(-2.7)\times10^{-2}$,
$B_{\mu\mu}^{R}=4.2$~$(1.5)\times10^{-2}$, and the central values of the other inputs, we get
\begin{equation}
A_{FB}(B\to K^{\ast} \mu^{+}\mu^{-})_{0 {\rm GeV}^2\leqslant q^2\leqslant 2 {\rm GeV}^2}=-0.02~(-0.05)\,
\end{equation}
in scenario S1~(S2). Compared with the SM prediction $\sim-0.10$, this observable could be enhanced about $80\%$~($50\%$) by the $Z^{\prime}$ contribution, implying that the scenario S1 with $\phi_{B_{s}}=-20.3^{\circ}\pm5.3^{\circ}$ fitted by UTfit collaboration~\cite{UTfit}
is favored by these decays. However, since $B\to X_s \mu^{+}\mu^{-}$ decay has already put strong constraint on the strength of the $Z^{\prime}$ couplings, the result for $A_{FB}(B\to K^{\ast}
\mu^{+}\mu^{-})_{0 {\rm GeV}^2\leqslant q^2\leqslant 2 {\rm GeV}^2}$ is still negative in S1~(S2),
being $1.5~(1.6)\sigma$ lower than the data $0.47^{+0.26}_{-0.32}$~\cite{:2009zv}. Such a situation
could also be seen from Fig.~\ref{FigDBA}, where the effects of $Z^{\prime}$ contribution induced by $B_{\mu\mu}^{L,R}$ on $d{\cal B}(B\to K^{(\ast)} \mu^+\mu^-)/d\hat{s}$ and $A_{FB}(B\to K^{\ast}
\mu^+\mu^-)$ are displayed.

With the bounded $Z^{\prime}$ couplings listed in Table~\ref{NPPara_value}, our predictions for ${\cal B}(B\to K^{(\ast)} \mu^+\mu^-)$ and $A_{FB}(B\to K^{\ast} \mu^{+}\mu^{-})$, both within the SM and in the $Z^{\prime}$ model, are given in Tables~\ref{tab_Kuu} and \ref{tab_Kstaruu}. We find that most of
the observables agree with the experimental data within errors. However, beside the discrepancy of $A_{FB}(B\to K^{\ast} \mu^+\mu^-)$ in the low $q^2$ bin, a similar problem is also observed in the high $q^2$ bin~($q^2\geqslant16{\rm GeV}^2$). Within the allowed parameter space shown in Fig.~\ref{FigBmumuLR}, the discrepancy for $A_{FB}(B\to K^{\ast}
\mu^+\mu^-)_{q^2\geqslant16{\rm GeV}^2}$ between the SM prediction and the experimental data is still difficult to be reconciled. In fact, if the $Z^{\prime}$ correction could not give a significant contribution after totally counteracting the SM contributions, the $A_{FB}(B\to K^{\ast} \mu^+\mu^-)$ problems in both the low and the high $q^2$ bins will persist within such a family non-universal $Z^{\prime}$ model.

\begin{table}[t]
 \begin{center}
 \caption{Predictions for ${\cal B}(B\to K \mu^+\mu^-)$ within the SM and the $Z^{\prime}$ model.}
 \label{tab_Kuu}
 \vspace{0.5cm}
 \doublerulesep 0.7pt \tabcolsep 0.1in
 \begin{tabular}{lccccccccccc} \hline \hline
  $q^2$ (${\rm GeV}^2$) & Exp.~\cite{:2009zv}     & SM            & S1           & S2
 \\\hline
 $[0,25]$     & $4.5\pm0.4$~\cite{HFAG} & $5.57\pm0.51$ &$5.74\pm0.84$ & $6.16\pm0.63$  \\
 $[1,6]$      & $1.36^{+0.24}_{-0.22}$  & $1.62\pm0.21$ &$1.64\pm0.29$ & $1.77\pm0.24$ \\
 $[0,2]$      & $0.81^{+0.19}_{-0.17}$  & $0.62\pm0.09$ &$0.61\pm0.12$ & $0.66\pm0.10$  \\
 $\geqslant16$        & $0.98^{+0.21}_{-0.19}$  & $0.84\pm0.03$ &$0.86\pm0.11$ & $0.92\pm0.07$ \\
  \hline \hline
 \end{tabular}
 \end{center}
 \end{table}

\begin{table}[t]
 \begin{center}
 \caption{Predictions for ${\cal B}(B\to K^{\ast} \mu^+\mu^-)$ and $A_{FB}(B\to K^{\ast} \mu^{+}\mu^{-})$ within the SM and the $Z^{\prime}$ model.}
 \label{tab_Kstaruu}
 \vspace{0.5cm}
 \doublerulesep 0.7pt \tabcolsep 0.1in
 \begin{tabular}{lccccccccccc} \hline \hline
           &$q^2$ (${\rm GeV}^2$) & Exp.~\cite{:2009zv}     & SM            & S1           & S2
 \\\hline
 ${\cal B}$&$[0,25]$    & $10.8^{+1.2}_{-1.1}$~\cite{HFAG}    & $11.3\pm0.2$  &$11.8\pm1.4$  & $12.6\pm0.9$  \\
           &$[1,6]$     & $1.49^{+0.47}_{-0.42}$  & $2.35\pm0.05$ &$2.45\pm0.31$ & $2.61\pm0.19$ \\
           &$[0,2]$     & $1.46^{+0.42}_{-0.37}$  & $1.23\pm0.06$ &$1.24\pm0.12$ & $1.26\pm0.08$  \\
          &$\geqslant16$& $2.04^{+0.32}_{-0.29}$  & $1.39\pm0.03$ &$1.45\pm0.18$ & $1.55\pm0.12$ \\
 \hline
 $A_{FB}$  &$[0,25]$    &---                      & $0.27\pm0.01$ &$0.25\pm0.05$ & $0.26\pm0.03$  \\
           &$[1,6]$     & $0.26^{+0.28}_{-0.31}$  & $0.07\pm0.01$ &$0.09\pm0.03$ & $0.09\pm0.02$ \\
           &$[0,2]$     & $0.47^{+0.26}_{-0.32}$  &$-0.10\pm0.01$ &$-0.05\pm0.02$& $-0.07\pm0.01$  \\
          &$\geqslant16$& $0.66^{+0.12}_{-0.16}$  & $0.34\pm0.01$ &$0.30\pm0.05$ & $0.32\pm0.03$ \\
  \hline \hline
 \end{tabular}
 \end{center}
 \end{table}

As a final comment, abandoning the constraints from $B\to X_s \mu^{+}\mu^{-}$ decay, we pursue the required strength of $Z^{\prime}$ couplings in order to make the sign of ${\rm Re}(C_7^{\rm eff}\bar{C}_{10}^{{\rm eff}\ast})$ flipped, while leaving ${\rm Re}(\bar{C}_9^{\rm eff}\bar{C}_{10}^{{\rm eff}\ast})$ unchanged. The conditions ${\rm Re}(C_7^{\rm eff}\bar{C}_{10}^{{\rm eff}\ast})<0$ and
${\rm Re}(\bar{C}_9^{\rm eff}\bar{C}_{10}^{{\rm eff}\ast})<0$ are equivalent to the following demands:
\begin{eqnarray}
{\rm Re}(C_{10}^{\prime})>0, \qquad {\rm Re}(C_9^{\prime})<0\, \qquad {\rm Im}(\triangle C_9^{\prime})>0\;.
\end{eqnarray}
From Eqs.~(\ref{NuVaC9}) and (\ref{NuVaC0}), and with the values of $|B_{sb}^L|$ and $\phi_{s}^L$ listed in Table~\ref{NPPara_value}, we get
\begin{eqnarray}\label{SpeCond}
B_{\mu\mu}^L<-0.271\;,\qquad 0.273+B_{\mu\mu}^L < B_{\mu\mu}^R <-0.268-B_{\mu\mu}^L\;,\qquad (\text{S1})\, \nonumber\\
B_{\mu\mu}^L<-0.287\;,\qquad 0.280+B_{\mu\mu}^L < B_{\mu\mu}^R <-0.275-B_{\mu\mu}^L\;,\qquad (\text{S2})\,
\end{eqnarray}
Taking $B_{\mu\mu}^L=-0.3$ and $B_{\mu\mu}^R=0.01$ in both S1 and S2, which satisfy Eq.~(\ref{SpeCond}) and are similar to the results given by Ref.~\cite{Chiang}, as shown in Fig.~\ref{FigDBA}~(c) $A_{FB}(B\to K^{\ast} \mu^+\mu^-)$ could be significantly enhanced to the experimental level. Unfortunately, such a region given by Eq.~(\ref{SpeCond}) is excluded by the constraint from $B\to X_s \mu^{+}\mu^{-}$ decay.

\section{Conclusion}

In conclusion, motivated by the large discrepancy for $A_{FB}(B\to K^{\ast} \mu^+\mu^-)$ in the low $q^2$ region, we have studied a family non-universal $Z^{\prime}$ model to pursue possible solution. With the constrained coupling $Z^{\prime}-\bar{s}b$ from $B_s-\bar{B}_s$ mixing and $B\to\pi K$ decays~\cite{Yang1}, we focus on the allowed regions for the couplings $Z^{\prime}-\mu^+\mu^-$ $B_{\mu\mu}^{L,R}$, which have already been strongly constrained by ${\cal B}(B\to X_s\mu^{+}\mu^{-})$ decay~\cite{Yang1} at high, low and full $q^2$ regions, with the further constraint from $A_{FB}(B\to K^{\ast} \mu^+\mu^-)_{0 {\rm GeV}^2\leqslant q^2\leqslant 2 {\rm GeV}^2}$. Within the allowed $Z^{\prime}$ couplings, we have investigated the effect of such a $Z^{\prime}$ model on the observables of $B\to K^{(\ast)}\mu^{+}\mu^{-}$ decays. Our main conclusions are summarized as:

\begin{itemize}
\item To account for the experimental data on $A_{FB}(B\to K^{\ast} \mu^+\mu^-)$, naively we get two interesting relations, $B_{\mu\mu}^{L}<B_{\mu\mu}^{R}$ and $B_{\mu\mu}^{L}+B_{\mu\mu}^{R}<0$. Furthermore, a larger $|B_{sb}^{L}|$ and a smaller $|\phi_s^{L}|$~(negative) are crucial to moderate the discrepancy of $A_{FB}(B\to K^{\ast} \mu^+\mu^-)_{0 {\rm GeV}^2\leqslant q^2\leqslant 2 {\rm GeV}^2}$ between the SM prediction and experimental data. Thus, scenario S1 is preferable to S2.

\item $A_{FB}(B\to K^{\ast} \mu^+\mu^-)_{0 {\rm GeV}^2\leqslant q^2\leqslant 2 {\rm GeV}^2}$ puts a strong constraint on the $Z^{\prime}$ couplings, $B_{\mu\mu}^{L,R}$. Including the constraints from $B\to X_s\mu^{+}\mu^{-}$ decay and $A_{FB}(B\to K^{\ast} \mu^+\mu^-)_{0 {\rm GeV}^2\leqslant q^2\leqslant 2 {\rm GeV}^2}$, we get $B_{\mu^{+}\mu^{-}}^{L}\sim-5(-2)\times10^{-2}$ and $B_{\mu^{+}\mu^{-}}^{R}\sim2(1)\times10^{-2}$ in S1~(S2).

\item Due to the severe constraints from $B\to X_s\mu^{+}\mu^{-}$ on the strength of the $Z^{\prime}$ contribution, the $A_{FB}(B\to K^{\ast} \mu^+\mu^-)$ problems in both low and high $q^2$ regions still persist. After including the $Z^{\prime}$ contribution, compared to the SM prediction, $A_{FB}(B\to K^{\ast} \mu^+\mu^-)_{0 {\rm GeV}^2\leqslant q^2\leqslant 2 {\rm GeV}^2}$ could be enhanced by an amount of about $80\%$~($50\%$), which is still about $1.5\sigma$~($1.6\sigma$) lower than the experimental data in S1~(S2).
\end{itemize}

Within such a family non-universal $Z^{\prime}$ model, although involving the same $Z^{\prime}-\bar{s}b$ coupling, these different processes also depend on different diagonal $Z^{\prime}$ couplings. For example, $B \to X_s \mu^+\mu^-$ and $B \to K^{(\prime)} \mu^+\mu^-$ depend on lepton diagonal coupling $Z^{\prime}-\mu^+\mu^-$, while $B \to K \pi$ decays on quark diagonal couplings $Z^{\prime}-u \bar{u}$ and $Z^{\prime}-d \bar{d}$. So, if one observable from the list of current anomalies~($A_{FB}$ at low dilepton mass, the $B_s$ mixing phase, the $K \pi$ puzzle) becomes SM-like, it will only affect the respective coupling, while leaving the others unchanged. For example, if the $B_s$ mixing phase becomes SM-like, then through adjusting the diagonal lepton and quark couplings, we can still find suitable parameter spaces to account for the other anomalies. If the $K \pi$ puzzle and/or the $A_{FB}$ at low dilepton mass become SM-like, it will only give more severe constraints on the diagonal lepton and quark couplings. To further constrain the model parameter spaces, it is therefore necessary to combine all of these processes at the same time and perform a global analysis, which is however beyond the scope of this paper but will be addressed in a forthcoming publication. With the upcoming LHC-b and proposed super-B experiments, the data on these processes is expected to be more precise~\cite{newdata}, which will then severely shrink or totally excluded the model.

\noindent{\it Note added}: During our work on the way, we note that a recent paper~\cite{Chiang} also pursues possible solutions within a family non-universal $Z^{\prime}$ model. In order to enhance the $Z^{\prime}$ contribution to the real part of $C_7^{\rm eff}C_{10}^{{\rm eff}\ast}$, they have assumed that $\phi_s^L=0$, which is obviously unsuitable due to the fact that a nonzero $\phi_s^L$ is needed to resolve the ``$\pi K$ puzzle'' and the ``$B_s-\bar{B}_s$ problem''~\cite{Yang1,Liu,Barger}. Furthermore, their result $B_{ll}^{L}\sim B_{ll}^{R}\sim{\cal O}(10^{-1})$ is also excluded by the constraint from $B\to X_s \mu^{+}\mu^{-}$ decays, in which $B_{ll}^{L}\sim B_{ll}^{R}\sim{\cal O}(10^{-2})$~\cite{Yang1}.

\section*{Acknowledgments}
X.~Q.~Li acknowledges support from the Alexander-von-Humboldt
Foundation. The work is supported by the National Science Foundation
under contract Nos.10675039 and 10735080.

\begin{appendix}

\section*{Appendix A: Theoretical input parameters}

For the CKM matrix elements, we adopt the fitting results from the UTfit collaboration~\cite{UTfit,UTfitCKM}
\begin{eqnarray}
&&\overline{\rho}=0.154\pm0.022\,(0.177\pm0.044), \nonumber\\
&&\overline{\eta}=0.342\pm0.014\,(0.360\pm0.031),\nonumber\\
&&|V_{td}/V_{ts}|=0.209\pm0.0075\,(0.206\pm0.012),\nonumber\\
&&|V_{cb}|=(4.13\pm0.05)\times10^{-2}\,((4.12\pm0.05)\times10^{-2}),
\end{eqnarray}
with $\overline{\rho}=\rho\,(1-\frac{\lambda^2}{2})$ and $\bar{\eta}=\eta\,(1-\frac{\lambda^2}{2})$. The values given in the brackets are the CKM parameters in presence of generic NP, and are used in our calculation when the $Z^{\prime}$ contributions are included.

As for the quark masses, we take~\cite{PDG08,PMass}
\begin{eqnarray}
 &&m_u=m_d=m_s=0, \quad m_c=1.61^{+0.08}_{-0.12}\,{\rm GeV},\nonumber\\
 &&m_b=4.79^{+0.19}_{-0.08}\,{\rm GeV}, \quad m_t=172.4\pm1.22\,{\rm GeV}.
\end{eqnarray}

\section*{Appendix B: Transition form factors from light-cone QCD sum rule}

In order to calculate the $B\to K^{(*)} \ell^+ \ell^-$ decay amplitude, we have to evaluate the $B \to K^{(*)}$ matrix elements of quark bilinear currents. They can be expressed in terms of ten form factors, which depend on the momentum transfer $q^2$ between the $B$ and the $K^{(*)}$ mesons~($q=p - k$)~\cite{Ball:2004rg}:
\begin{eqnarray}
\langle \bar{K}(k)|\bar{s} \gamma_{\mu} (1-\gamma_{5}) b|\bar{B}(p)\rangle&=&f_+(q^2)\Big[(2p-q)_{\mu}-\frac{m_{B}^{2}-m_{K}^{2}}{q^2}q_{\mu}\Big]
+\frac{m_{B}^{2}-m_{K}^{2}}{q^2}f_{0}(q^2)q_{\mu}\,,\\
\langle \bar{K}(k)|\bar{s}\sigma_{\mu\nu}q^{\nu}(1+\gamma_{5} )b|\bar{B}(p)\rangle&=&i\Big[(2p-q)_{\mu}q^2-q_{\mu}(m_B^2-m_K^2)\Big]\frac{f_{T}(q^2)}{m_B+m_K}\, ,
\end{eqnarray}
with $f_+(0)=f_0(0)$,
\begin{eqnarray}\label{eq:SLFF}
\langle \bar K^*(k) | \bar s\gamma_\mu(1-\gamma_5) b | \bar
B(p)\rangle  &=& -i \epsilon^*_\mu (m_B+m_{K^*}) A_1(q^2) + i
(2p-q)_\mu (\epsilon^* \cdot q)\,
\frac{A_2(q^2)}{m_B+m_{K^*}}\, \nonumber \\
& & + i q_\mu (\epsilon^* \cdot q) \, \frac{2m_{K^*}}{q^2}\,
\Big[A_3(q^2)-A_0(q^2)\Big] \, \nonumber \\
& & + \epsilon_{\mu\nu\rho\sigma}\epsilon^{*\nu} p^\rho k^\sigma\,
\frac{2V(q^2)}{m_B+m_{K^*}}\,,
\end{eqnarray}
with $A_3(q^2) = \frac{m_B+m_{K^*}}{2m_{K^*}}\, A_1(q^2) -
\frac{m_B-m_{K^*}}{2m_{K^*}}\, A_2(q^2)$ and $A_0(0) =  A_3(0)$,
\begin{eqnarray}\label{eq:pengFF}
\langle \bar K^*(k) | \bar s \sigma_{\mu\nu} q^\nu (1+\gamma_5) b |
\bar B(p)\rangle &=& i\epsilon_{\mu\nu\rho\sigma} \epsilon^{*\nu}
p^\rho k^\sigma \, 2 T_1(q^2)\, \nonumber\\
& & + T_2(q^2) \Big[\epsilon^*_\mu (m_B^2-m_{K^*}^2)
- (\epsilon^* \cdot q) \,(2p-q)_\mu \Big] \, \nonumber\\
& & + T_3(q^2) (\epsilon^* \cdot q) \left[q_\mu -
\frac{q^2}{m_B^2-m_{K^*}^2}\, (2p-q)_\mu \right]\,,
\end{eqnarray}
with $T_1(0) = T_2(0)$. $\epsilon_\mu$ is the polarization vector of the $K^*$ meson. The physical range in $s=q^2$ extends from $s_{\rm min} = 0$ to $s_{\rm max} =(m_B-m_{K^{(\ast)}})^2$.

These transition form factors have been updated recently within the light-cone QCD sum rule approach~\cite{Ball:2004rg}. For the $q^2$ dependence of the form factors, they can be parameterized in terms of simple formulae with two or three parameters. The form factors $V$, $A_0$ and $T_1$ are parameterized by
\begin{eqnarray}
F(s)=\frac{r_1}{1-s/m^2_{R}}+\frac{r_2}{1-s/m^2_{\rm fit}}.
\label{r12mRfit}
\end{eqnarray}
For the form factors $A_2$, $\tilde{T}_3$, $f_+$ and $f_T$, it is more appropriate to expand to the second order around the pole, yielding
\begin{eqnarray}
F(s)=\frac{r_1}{1-s/m^2}+\frac{r_2}{(1-s/m)^2}\,, \label{r12mfit}
\end{eqnarray}
where $m=m_{\rm fit}$ for $A_2$ and $\tilde{T}_3$, and $m=m_{R}$ for $f_+$ and $f_T$. The fit formula for $A_1$, $T_2$ and $f_0$ is
\begin{eqnarray}
F(s)=\frac{r_2}{1-s/m^2_{\rm fit}}.\label{r2mfit}
\end{eqnarray}
The form factor $T_3$ can be obtained through the relation $T_3(s)=\frac{1-m_{K^*}}{s}\big[\tilde{T}_3(s)-T_2(s)\big]$. All the relevant fitting parameters for these form factors are taken from Ref.~\cite{Ball:2004rg} and are recollected in Table~\ref{FFfit}.

\begin{table}[t]
\begin{center}
\caption{\label{FFfit} Fit parameters for $B\to K^{(*)}$ transition form
factors~\cite{Ball:2004rg}.}
\vspace{0.3cm}
\begin{tabular}{crrrrrl}\hline\hline
                &$F(0)$ &$r_1$   &$m_R^2$ &$r_2$   &$m^2_{\rm fit}$& \\ \hline
$f_+^{B\to K}$   &0.331  &0.162  &$5.41^2$&0.173   &         &Eq.~(\ref{r12mfit})   \\\hline
$f_0^{B\to K}$   &0.331  &       &        &0.330   &$37.46$  &Eq.~(\ref{r2mfit})   \\\hline
$f_T^{B\to K}$   &0.358  &0.161  &$5.41^2$&0.198   &         &Eq.~(\ref{r12mfit})   \\\hline
$V^{B\to K^*}$  &$0.411$&$0.923$ &$5.32^2$&$-0.511$&$49.40$&Eq.~(\ref{r12mRfit})\\\hline
$A_0^{B\to K^*}$&$0.374$&$1.364$ &$5.28^2$&$-0.990$&$36.78$&Eq.~(\ref{r12mRfit})\\\hline
$A_1^{B\to K^*}$&$0.292$&$$      &$$&$0.290$&$40.38$&Eq.~(\ref{r2mfit})\\\hline
$A_2^{B\to K^*}$&$0.259$&$-0.084$&$$&$0.342$&$52.00$&Eq.~(\ref{r12mfit})\\\hline
$T_1^{B\to K^*}$&$0.333$&$0.823$ &$5.32^2$&$-0.491$&$46.31$&Eq.~(\ref{r12mRfit})\\\hline
$T_2^{B\to K^*}$&$0.333$&$$      &$$&$0.333$&$41.41$&Eq.~(\ref{r2mfit})\\\hline
$\tilde{T}_3^{B\to K^*}$&$0.333$ &$-0.036$&$$&$0.368$&$48.10$&Eq.~(\ref{r12mfit})\\
\hline\hline
\end{tabular}
\end{center}
\end{table}

\end{appendix}

\end{document}